\documentclass[preprintnumbers,unsortedaddress,superscriptaddress,notitlepage]{revtex4}
\usepackage{amsfonts}
\usepackage{amsmath}
\usepackage{hyperref}
\usepackage{amssymb}
\usepackage[english]{babel}
\usepackage{graphicx}
\usepackage{epsfig}
\usepackage{bm}
\usepackage{longtable}
\usepackage{verbatim}
\usepackage{longtable}
\usepackage{cleveref}

\crefname{equation}{Eq.}{Eqs.}
\newcommand{\mathsym}[1]{{}}

\newcommand{\emp}{\begin{equation}}
\newcommand{\fin}{\end{equation}}
\newcommand{\empn}{\begin{equation*}}
\newcommand{\finn}{\end{equation*}}

\newcommand{\bea}{\begin{eqnarray}}
\newcommand{\eea}{\end{eqnarray}}
\newcommand{\eger}{\begin{gather}}
\newcommand{\fger}{\end{gather}}
\newcommand{\egn}{\begin{gather*}}
\newcommand{\fgn}{\end{gather*}}
\newcommand{\bit}{\begin{itemize}}
\newcommand{\eit}{\end{itemize}}

\input{tcilatex}
\begin{document}

\title{Higgs boson coupling to a new strongly interacting sector}
\author{A. E. C\'arcamo Hern\'andez}
\email{antonio.carcamo@usm.cl}
\affiliation{{\small Universidad T\'ecnica Federico Santa Mar\'{\i}a and Centro Cient%
\'{\i}fico-Tecnol\'ogico de Valpara\'{\i}so}\\
Casilla 110-V, Valpara\'{\i}so, Chile}
\author{Claudio Dib}
\email{claudio.dib@usm.cl}
\affiliation{{\small Universidad T\'ecnica Federico Santa Mar\'{\i}a and Centro Cient%
\'{\i}fico-Tecnol\'ogico de Valpara\'{\i}so}\\
Casilla 110-V, Valpara\'{\i}so, Chile}
\author{Alfonso Zerwekh}
\email{alfonso.zerwekh@usm.cl}
\affiliation{{\small Universidad T\'ecnica Federico Santa Mar\'{\i}a and Centro Cient%
\'{\i}fico-Tecnol\'ogico de Valpara\'{\i}so}\\
Casilla 110-V, Valpara\'{\i}so, Chile}

\begin{abstract}
In the framework of strongly interacting dynamics for electroweak symmetry
breaking, heavy composite particles may arise and cause observable effects,
as they should couple strongly to the resulting Higgs boson and affect the
signals that appear at one loop level. Here we study this expected behavior,
contrasting it with current experimental knowledge. We work in a simple and
generic scenario where the lowest lying composite states are the Higgs
scalar doublet and a massive vector triplet. We use an effective chiral
Lagrangian to describe the theory below the compositeness scale $\Lambda$,
assumed to be $4\pi v \simeq~3$~TeV. The effective theory contains the
Standard Model spectrum and the extra composites. We determine the
constraints on this scenario imposed by our current knowledge of the $Zb\bar{%
b}$ vertex, the $T$ and $S$ oblique parameters, and the recently measured
Higgs mass and its diphoton decay rate. We found that the $T$ and $S$
parameters as well as the Higgs diphoton decay do not provide important
constraints on the model. In contrast, the constraints arising from the $Zb%
\bar{b}$ vertex and from the Higgs mass at $126$ GeV are fulfilled only if
the heavy vector resonances do not couple strongly with quarks, and at the
same time the Higgs boson has a moderate but not too strong coupling to the
heavy composite resonances.
\end{abstract}

\maketitle

%\maketitle
%\maketitle

%\PACS{
%  {12.60.Rc}{Composite models}   \and
%{12.60.Cn}{Extensions of electroweak gauge sector} \and
%{12.60.Fr}{Extensions of electroweak Higgs sector.} 
%  } % end of PACS codes
%} %end of abstract
%
%
%\authorrunning{A. C\'arcamo, C. Dib, A. Zerwekh} 
%\titlerunning{Composite Higgs Coupling to a New Strong Sector}

\section{Introduction.}

\label{intro} The recent discovery of the Higgs boson at the LHC \cite%
{atlashiggs,cmshiggs,newtevatron,CMS-PAS-HIG-12-020} provides the
opportunity to directly explore the mechanism of Electroweak Symmetry
breaking. While this remarkable achievement implies severe constrains on
many proposed extensions of the Standard Model, an additional sector beyond
our current knowledge is still needed in order to explain the dynamical
origin of the electroweak scale and its stability \cite{pdg}. A specific
question in this context is whether this new sector is weakly or strongly
interacting \cite{Contino:2009ez}. In the latter case, the Higgs boson is
viewed as a composite state which must be accompanied by a plethora of new
heavy composite particles \cite{Grojean:2009fd,Contino:2010rs,Panico:2015jxa}%
. In general, it is expected that the lightest states produced by the strong
dynamics would correspond to spin-0 and spin-1 particles \cite%
{Grojean:2009fd,Contino:2010rs,Panico:2015jxa}. In these models the
lightness of the Higgs can be explained in two different ways. One way is to
consider the Higgs boson as a pseudo-Goldstone boson that appears after the
breakdown of a suitable global symmetry \cite%
{Contino:2010rs,Panico:2015jxa,Barbieri:2007,Csaki:2008zd,Contino:2012,Pomarol:2012,Pappadopulo:2013vca,Montull:2013mla}%
. A second way is to consider the Higgs boson as the modulus of an effective
SU(2) doublet, where its lightness is due to particularities of the dynamics
of the underlying theory \cite%
{Bardeen:1989ds,ggpr,Zerwekh:2006,Zerwekh:2010,Carcamo:2010,Carcamo:ggtoVV,Carcamo:2011,Burdman:2011,Carcamo:2012,Bellazini:2012,Contino:2013kra,Diaz:2013tfa,Castillo-Felisola:2013jua,Hernandez:2013zho,Carcamo-Hernandez:2013ypa,Carcamo-Hernandez:2015xka,Pappadopulo:2014qza,Foadi:2007ue,Ryttov:2008xe,Sannino:2009za,Belyaev:2013ida,Hapola:2011sd,Foadi:2012bb}%
. For instance, there are evidences that quasi-conformal strong interacting
theories such as Walking Technicolor may provide a light composite scalar 
\cite%
{Foadi:2007ue,Ryttov:2008xe,Sannino:2009za,Belyaev:2013ida,Hapola:2011sd,Foadi:2012bb}%
. It has also been shown that, in the effective low energy theory, the
composite scalar may develop a potential that reproduces the standard Higgs
sector \cite{Bardeen:1989ds}. In this scheme, the Electroweak Symmetry
breaking is effectively described by a non zero vacuum expectation value of
the scalar arising from the potential, just as in the Standard Model.
However, additional composite particles, like vector resonances, may also be
expected to appear in the spectrum. In such a scenario, the vector sector
can be extended by introducing a composite triplet of heavy vector
resonances. This is the path we follow in this paper.

In general, a composite Higgs boson is theoretically attractive because the
underlying strong dynamics provides a comprehensive and natural explanation
for the origin of the Fermi scale \cite%
{Grojean:2009fd,Contino:2010rs,Panico:2015jxa}. However, the presence of the
already mentioned additional composite states may, in principle, produce
phenomenological problems. For instance one could expect that, at one loop
level, they may produce sensible corrections to observables involving the
Higgs boson. Consequently, an interesting quantity which can eventually
reveal the influence of additional states is $\Gamma (h\rightarrow \gamma
\gamma )$. In a previous work one of us studied this decay channel in a
simple model with vector resonances and found it is in general agreement
with current experimental measurements in the limit where the Higgs boson is
weakly coupled to the new resonances \cite{Castillo-Felisola:2013jua}. In
this work, we want to go further and use this channel to investigate if the
experimental results still allow for a Higgs boson strongly coupled to the
new $SU\left( 2\right) _{L}$ triplet of heavy vectors\ resonances. In order
to be concrete and predictive, we describe the new sector by means of an
effective model with minimal particle content. We use an effective chiral
Lagrangian to describe the theory below the cutoff scale, assumed to be $%
\Lambda =4\pi v\sim 3$ TeV, which contains the Standard Model spectrum and
the extra composites.

The content of this paper goes as follows. In section \ref{Model} we
introduce our effective Lagrangian that describes the spectrum of the
theory. In section \ref{Zbb} we discuss the implications of the $Zb\bar{b}$
constraint in our model. In section \ref{TandS}, we determine the $T$ and $S$
oblique parameters in our model and discuss the implications of our model
for electroweak precision tests. Section \ref{Higgsdiphotonrate} deals with
the constraints arising from the Higgs diphoton decay rate and the
requirement of having a composite scalar mass of $126$ GeV. In section \ref%
{Decaychannels} we describe the different decay channels of the heavy vector
resonances. Finally in Section \ref{conclusions}, we state our conclusions.

%%%%%%%%%%%%%%%%%%%%%%%%%%%
%%% SECTION 2
%%%%%%%%%%%%%%%%%%%%%%%%%%%

\section{Lagrangian for a Higgs doublet and heavy vector triplet.}

\label{Model} %%%%%%%%%%%%%%%%%%%%%%%%%%%%
% SECTION 3
%%%%%%%%%%%%%%%%%%%%%%%%%%%%

To determine whether the current experimental data still allows for Higgs
boson strongly coupled to a composite sector, we start by formulating our
strongly coupled sector by means of an effective theory based on the gauge
group $SU\left( 2\right) _{L}\times U\left( 1\right) _{Y}$ with a $SU\left(
2\right) _{L}$ triplet of heavy vectors in addition to the SM fields. The
gauge symmetry of the Standard Model is broken when the electrically neutral
component of the scalar doublet $\Phi $ acquires a vacuum expectation value.
The model Lagrangian can be written as: 
\begin{eqnarray}
\tciLaplace &=&-\frac{1}{2g^{2}}\left\langle W_{\mu \nu }W^{\mu \nu
}\right\rangle -\frac{1}{2g^{\prime 2}}\left\langle B_{\mu \nu }B^{\mu \nu
}\right\rangle -\frac{1}{2g_{\rho }^{2}}\left\langle \rho _{\mu \nu }\rho
^{\mu \nu }\right\rangle  \notag \\
&&+f^{2}\left\langle \rho _{\mu }\rho ^{\mu }\right\rangle +\left( D_{\mu
}\Phi \right) ^{\dag }D^{\mu }\Phi -\frac{\lambda }{4}\left( \Phi ^{\dag
}\Phi \right) ^{2}+\mu ^{2}\Phi ^{\dag }\Phi  \notag \\
&&+\alpha _{1}\func{Re}\left( \Phi ^{\dag }\rho ^{\mu }D_{\mu }\Phi \right)
+\alpha _{2}\func{Im}\left( \Phi ^{\dag }\rho ^{\mu }D_{\mu }\Phi \right) 
\notag \\
&&+\beta _{1}\left\langle \rho _{\mu }\rho ^{\mu }\right\rangle \Phi ^{\dag
}\Phi +\beta _{2}\Phi ^{\dag }\rho _{\mu }\rho ^{\mu }\Phi  \notag \\
&&+i\kappa _{0}\left\langle W^{\mu \nu }\left[ \rho _{\mu },\rho _{\nu }%
\right] \right\rangle  \notag \\
&&+i\kappa _{1}\left\langle \rho ^{\mu \nu }\left[ \rho _{\mu },\rho _{\nu }%
\right] \right\rangle +\kappa _{2}\left\langle \left[ \rho ^{\mu },\rho
^{\nu }\right] \left[ \rho _{\mu },\rho _{\nu }\right] \right\rangle  \notag
\\
&&+\frac{ig_{\rho q\overline{q}}}{2}\left( \overline{q}_{iL}\gamma ^{\mu
}q_{iL}\rho _{\mu }^{3}+\sqrt{2}V_{ij}\overline{u}_{iL}\gamma ^{\mu
}d_{jL}\rho _{\mu }^{+}+h.c\right) +...,  \label{ModelLagrangian}
\end{eqnarray}%
where all parameters appearing in Eq. (\ref{ModelLagrangian}) are
dimensionless with the exception of $f$ and $\mu $, which have mass
dimension. Furthermore, the omitted terms ($...$) are the extra terms such
as the couplings of the SM gauge bosons with SM fermions as well as the
couplings of the heavy vectors with SM leptons. Here $\left\langle
{}\right\rangle $ denotes the trace over the $2\times 2$ matrices and the
scalar doublet $\Phi $ contains the SM Higgs boson and the SM
would-be-Goldstone bosons. The SM Higgs doublet is given as usual by:

\begin{equation}
\Phi =\left( 
\begin{array}{c}
G^{+} \\ 
\frac{1}{\sqrt{2}}\left( v+h+i\eta \right)%
\end{array}%
\right) =\left( 
\begin{array}{c}
\frac{1}{\sqrt{2}}\left( \omega +i\xi \right) \\ 
\frac{1}{\sqrt{2}}\left( v+h+i\eta \right)%
\end{array}%
\right) ,
\end{equation}
where the effective fields $h$, $\eta$, $\omega$ and $\xi$ have zero vacuum
expectation values.

The covariant derivate acting on the scalar doublet $\Phi $\ can be written as follows: 
\begin{equation}
D_{\mu }\Phi =\partial _{\mu }\Phi -\frac{i}{2}gW_{\mu }^{a}\tau ^{a}\Phi -%
\frac{i}{2}g^{\prime }B_{\mu }\Phi .
\end{equation}

Furthermore, the tensor $\rho _{\mu \nu }=D_{\mu }\rho _{\nu }-D_{\nu }\rho
_{\mu }$ is written in terms of a covariant derivative of the field $%
\rho_\mu $ as follows: 
\begin{equation}
D_{\mu }\rho _{\nu }=\partial _{\mu }\rho _{\nu }-\frac{i}{2}\left[ W_{\mu
},\rho _{\nu }\right] ,
\end{equation}%
where $\rho _{\mu }=g_{\rho }\rho _{\mu }^{a}\frac{\tau ^{a}}{2}$ is a $%
SU\left( 2\right) _{L}$ triplet of composite vector resonances, neutral
under hypercharge, formed due to the underlying strong dynamics.

In view of the large number of parameters of the model [c.f Eq. (\ref%
{ModelLagrangian})] and in order to make definite predictions, we assume
that the interactions of the heavy vectors $\rho _{\mu }$ with themselves as
well as with the SM gauge bosons have the same Lorentz structure as the
interactions among the SM gauge bosons. Consequently, the heavy vectors will
correspond to the gauge vectors of a hidden local symmetry, and the
aforementioned assumption leads to the following constraint: 
\begin{equation}
\kappa _{1}=2\kappa _{0}=2\kappa _{2}=\frac{1}{g_{\rho }^{2}}.
\end{equation}

When the Higgs boson acquires a vacuum expectation value $\left\langle \Phi
\right\rangle =\frac{1}{\sqrt{2}}\left( 0,v\right) ^{T}$, from Eq. (\ref{ModelLagrangian}) it follows that the squared mass matrices for the neutral
and charged gauge bosons are given by:

\begin{eqnarray}
M_{N}^{2} &=&\left( 
\begin{array}{ccc}
\frac{g^{\prime 2}}{4} & -\frac{gg^{\prime }}{4} & \alpha _{2}\frac{%
g^{\prime }g_{\rho }}{8} \\ 
-\frac{gg^{\prime }}{4} & \frac{g^{2}}{4} & -\alpha _{2}\frac{gg_{\rho }}{8}
\\ 
\alpha _{2}\frac{g^{\prime }g_{\rho }}{8} & -\alpha _{2}\frac{gg_{\rho }}{8}
& \left( \beta _{1}+\frac{\beta _{2}}{2}\right) \frac{g_{\rho }^{2}}{2}%
+g_{\rho }^{2}\frac{f^{2}}{v^{2}}%
\end{array}%
\right) v^{2},  \notag \\
M_{C}^{2} &=&\left( 
\begin{array}{cc}
\frac{g^{2}}{4} & -\frac{gg_{\rho }}{8}\left( \alpha _{2}+i\alpha _{1}\right)
\\ 
-\frac{gg_{\rho }}{8}\left( \alpha _{2}-i\alpha _{1}\right) & \left( \beta
_{1}+\frac{\beta _{2}}{2}\right) \frac{g_{\rho }^{2}}{2}+g_{\rho }^{2}\frac{%
f^{2}}{v^{2}}%
\end{array}%
\right) v^{2}.  \notag \\
&&  \label{Gaugebosonmasses}
\end{eqnarray}

Assuming $g_{\rho }f \gg 100$ GeV %$g_{\rho }>>g,g^{\prime }$ 
and considering all dimensionless couplings here of the same order of
magnitude, the following expressions for the gauge boson masses are
obtained: 
\begin{eqnarray}
M_{A} &=&0,\quad M_{Z}\simeq \frac{\sqrt{g^{2}+g^{\prime 2}}}{2}v,\quad
M_{W}\simeq \frac{gv}{2},  \notag \\
M_{\rho ^{0}} &\simeq & M_{\rho ^{\pm}} \simeq\sqrt{\beta _{1}+\frac{\beta
_{2}}{2}+\frac{2f^{2}}{v^{2}}}\frac{g_{\rho }v}{\sqrt{2}}.
\label{gaugebosonsmasses}
\end{eqnarray}

\begin{figure*}[tbh]
\begin{center}
\resizebox{0.3\textwidth}{!}{
\includegraphics{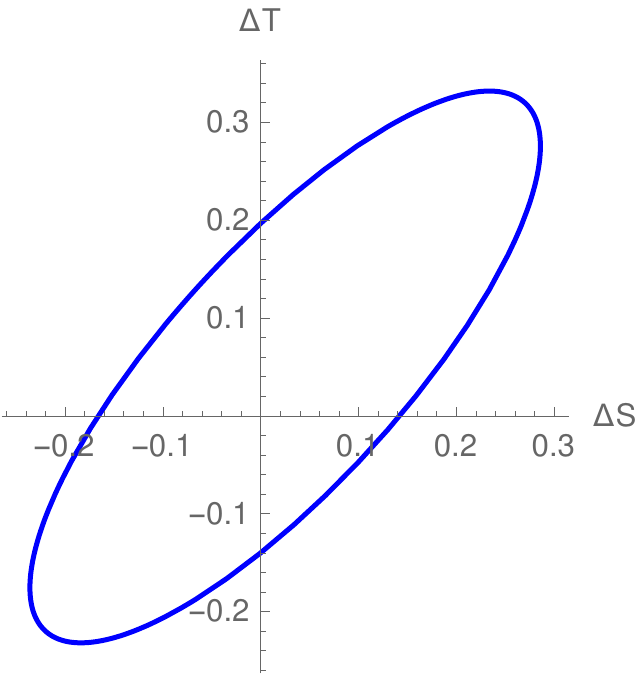}}
\end{center}
\caption{The interior of the ellipse in the $\Delta S-\Delta T$ plane is the
experimentally allowed region at $95\%$CL from Ref. \protect\cite{GFitter}.
The origin $\Delta S=\Delta T=0$ corresponds to the Standard Model value,
with $m_{h}=126$~GeV and $m_{t}=176$~GeV.}
\label{TandSexp}
\end{figure*}

\section{Constraints from the $Zb\bar{b}$ vertex}

\label{Zbb}

In this section we compute the one-loop correction to the $Zb\bar{b}$ vertex
in our model, in order to set an upper bound on the direct coupling $g_{\rho
q\overline{q}}$ of the heavy vectors with quarks. In the SM, the one-loop
diagram containing the top quark induces a sizeable modification, $A_{bb}$,
of the $Zb\bar{b}$ vertex from its tree level value: 
\begin{equation}
\left( -\frac{1}{2}+\frac{\sin ^{2}\theta _{W}}{3}+A_{bb}\right) \frac{g}{%
\cos \theta _{W}}Z_{\mu }\overline{b}_{L}\gamma ^{\mu }b_{L},
\end{equation}
where, in the large $m_{t}$ limit, the SM value for $A_{bb}$ is \cite%
{Barbieri:2007}:

\begin{equation}
A_{bb}^{\left( SM\right) }\simeq \frac{m_{t}^{2}}{16\pi ^{2}v^{2}} .
\end{equation}

In addition, the one-loop diagram containing a charged heavy vector
resonance $\rho _{\mu }^{\pm }$, gives the following contribution to $A_{bb}$%
, up to corrections of order $\frac{m_{t}^{2}}{M_{\rho }^{2}}$: 
\begin{equation}
\delta A_{bb}\simeq -\frac{41g_{\rho q\overline{q}}^{2}}{768\pi ^{2}}.
\end{equation}

Thus, the $A_{bb}$ parameter in our model is:

\begin{equation}
A_{bb} \simeq A_{bb}^{\left( SM\right) } \left(1-g_{\rho q\overline{q}}^{2}%
\frac{41}{48}\frac{v^{2}}{m_{t}^{2}}\right) .
\end{equation}

The experimental value of this quantity is $A_{bb}^{\left( \exp \right)
}=0.923\pm 0.020$ whereas its SM value is $A_{bb}^{\left( SM\right) }=0.9347$
\cite{pdg}. From these values we can extract bounds for $g_{\rho q\overline{q%
}}$, the direct coupling of the heavy vector resonances with quarks: the
requirement of having $A_{bb}$ consistent up to $1\sigma $ with the
experimental data yields the upper bound $g_{\rho q\overline{q}}\lesssim
0.14 $, while a consistency within $3\sigma$ gives the looser bound $g_{\rho
q\overline{q}}\lesssim~0.21$.

\section{Constraints from the T and S parameters}

\label{TandS}

The inclusion of the extra composite particles also modifies the oblique
corrections of the SM, the values of which have been extracted from high
precision experiments. Consequently, the validity of our model depends on
the condition that the extra particles do not contradict those experimental
results. These oblique corrections are parametrized in terms of the two well
known quantities $T$ and $S$. The $T$ parameter is defined as \cite{Peskin:1991sw,Peskin:1991sw2,epsilon-approach,epsilon-approach2,Barbieri:2004,Barbieri-book}%
: 
\begin{equation}
T=\frac{\Pi _{33}\left( 0\right) -\Pi _{11}\left( 0\right) } {%
M_{W}^{2}\alpha _{em}\left( m_{Z}\right) },  \label{T}
\end{equation}
where $\Pi _{33}\left( 0\right) $ and $\Pi _{11}\left( 0\right) $ are the
vacuum polarization amplitudes at $q^2 =0$ for the propagators of the gauge
bosons $\tilde{W}_{\mu }^{3}$ and $\tilde{W}_{\mu }^{1}$, respectively. Let
us note, as stressed by \cite{Barbieri:2004}, that the gauge bosons $\tilde{W%
}_{\mu }^{a}$ ($a=1,2,3$) are linear combinations of the $W_{\mu }^{a}$ ($%
a=1,2,3$) SM gauge fields and the heavy vector resonances $\rho_{\mu }^{a}$ (%
$a=1,2,3$) since these fields have direct couplings with SM fermions. These
linear combinations are defined in such a way that $\tilde{W}_{\mu }^{a}$ as
well as $B^0$ are the only spin-1 fields (apart from the gluons) having
gauge interactions with SM fermions. Consequently the fields $\tilde{W}_{\mu
}^{a}$ ($a=1,2,3$) are defined as: 
\begin{equation}
\tilde{W}_{\mu }^{a}=\frac{g}{\sqrt{g^2+g^2_{\rho q\overline{q}}}}W^a_{\mu}+%
\frac{g_{\rho q\overline{q}}}{\sqrt{g^2+g^2_{\rho q\overline{q}}}}%
\rho^a_{\mu},\hspace{1cm}a=1,2,3.
\end{equation}
\begin{figure*}[tbh]
\resizebox{0.95\textwidth}{!}{
\includegraphics{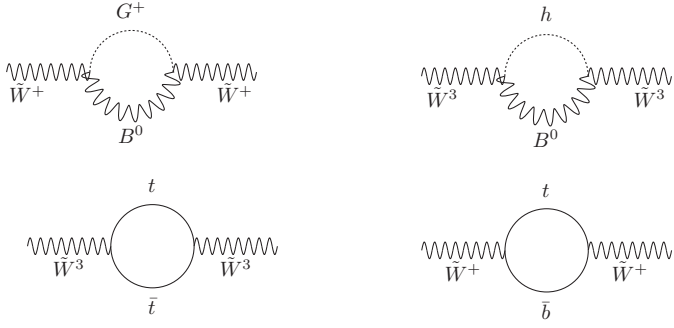}}\vspace{-16.5cm}
\caption{One loop Feynman diagrams contributing to the $T$ parameter.}
\label{figT}
\end{figure*}

\begin{figure*}[tbh]%\resizebox{17cm}{6cm}
\resizebox{0.95\textwidth}{!}{\includegraphics{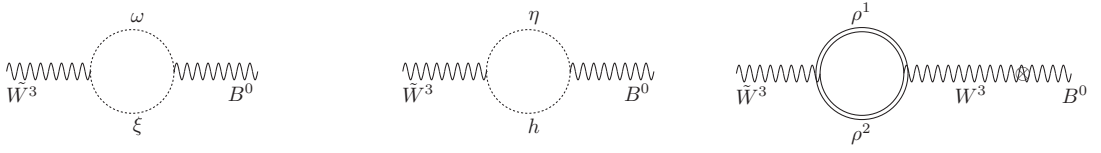}}\vspace{-19cm}
\caption{One loop Feynman diagrams contributing to the $S$ parameter. The
first and second diagrams correspond to the SM contributions, whereas the
third one is the contribution due to the heavy vector resonances.}
\label{figS}
\end{figure*}
In turn, the $S$ parameter is defined as \cite%
{Peskin:1991sw,Peskin:1991sw2,epsilon-approach,epsilon-approach2,Barbieri:2004,Barbieri-book}%
: 
\begin{equation}
S=\frac{4\sin ^{2}\theta _{W}}{\alpha _{em}\left( m_{Z}\right) }\frac{g}{%
g^{\prime }}\frac{d}{dq^{2}}\Pi _{30}\left( q^{2}\right) \biggl|_{q^{2}=0},
\end{equation}%
where $\Pi _{30}\left( q^{2}\right) $ is the vacuum polarization amplitude
for the propagator mixing of $\tilde{W}_{\mu }^{3}$ and $B_{\mu }$. The most
important Feynman diagrams contributing to the $T$ and $S$ parameters are
shown in Figures \ref{figT} and \ref{figS}. We computed these oblique $T$
and $S$ parameters in the Landau gauge for the SM gauge bosons and
would-be-Goldstone bosons, where the global $SU(2)_{L}\times U(1)_{Y}$
symmetry is preserved. We can separate the contributions to $T$ and $S$ from
the SM and extra physics as $T=T_{SM}+\Delta T$ and $S=S_{SM}+\Delta S$,
where 
\begin{eqnarray}
T_{SM} &=&-\frac{3}{16\pi \cos ^{2}\theta _{W}}\ln \left( \frac{m_{h}^{2}}{%
m_{W}^{2}}\right) +\frac{3m_{t}^{2}}{32\pi ^{2}\alpha _{em}\left(
m_{Z}\right) v^{2}},\hspace{0.2cm}  \notag \\
S_{SM} &=&\frac{1}{12\pi }\ln \left( \frac{m_{h}^{2}}{m_{W}^{2}}\right) +%
\frac{1}{2\pi }\left[ 3-\frac{1}{3}\ln \left( \frac{m_{t}^{2}}{m_{b}^{2}}%
\right) \right] ,
\end{eqnarray}%
%
%
%
%
%
%
%
%are the contributions within the SM, 
while $\Delta T$ and $\Delta S$ contain all the contributions involving the
extra particles. 
%The one loop diagrams that contribute to the $T$ parameter in the SM include the hypercharge gauge boson $B_{\mu }^{0}$, since the $g^{\prime }$ coupling is one of the sources of custodial symmetry breaking. The other source in the SM comes from the difference between up- and down-type quark Yukawa couplings, which is essentially dominated by the large value of the top quark mass. It is noteworthy that the gauge boson contribution to the $T$ parameter does not depend explicity on the hypercharge coupling $g^{\prime}$, since the gauge sector contributions to the $\Pi _{33}\left( 0\right) $ and $\Pi _{11}\left( 0\right)$ vacuum polarization amplitudes as well as the $\alpha _{em}$ coupling are directly proportional to $g^{\prime}$, so that the $g^{\prime}$ factor is cancelled.
%%
%%
%Regarding the heavy composite spin-1 resonances, we use the unitary gauge for their propagators since the Lagrangian given in Eq. (\ref{Leff}) does not include the Goldstone bosons associated to the longitudinal components of these heavy spin-1 resonances. %From the Feynman diagrams shown in Figure \ref{figT}, it follows that the $\widehat{T}$ parameter isgiven by  in Fig. 
\begin{figure*}[tbh]
\begin{center}
\vspace{1cm} \resizebox{0.6\textwidth}{!}{%
\includegraphics{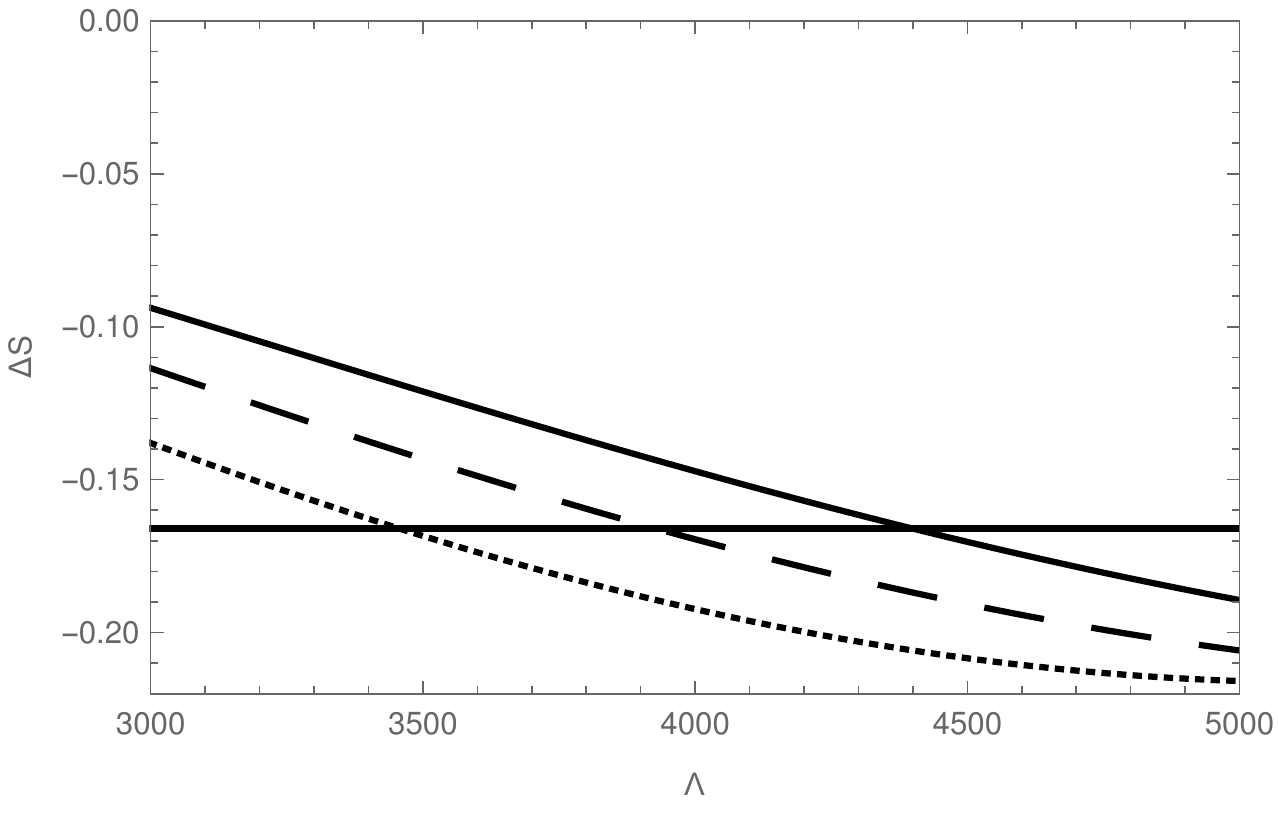}}
\caption{The $\Delta S$ parameter as a function of the cutoff $\Lambda $ for
different values of the heavy vector masses $M_{\protect\rho }$. Here we set 
$\protect\alpha_2=g_{\protect\rho }=1$ and $g_{\protect\rho q\overline{q}%
}=0.14$. The curves from top to bottom correspond to $M_{\protect\rho }=2.8$
TeV, $M_{\protect\rho }=2.5$ TeV and $M_{\protect\rho }=2.2$ TeV,
respectively. The horizontal line corresponds to the minimum experimental
value of the $\Delta S$ parameter at $95\%$CL for $\Delta T=0$.}
\label{DeltaSvsLambda}
\end{center}
%\par
%The blue, red and green curves correspond to $M_{\protect\rho }=2.2$ TeV, $M_{\protect\rho }=2.5$ TeV and $M_{\protect\rho }=2.8$ TeV, respectively
\end{figure*}
%\newline
The absence of quadratic divergences in the $T$ parameter requires that $%
\alpha _{1}=0$ in Eq.~(\ref{ModelLagrangian}). In that case the dominant
one-loop contribution to $\Delta T$ and $\Delta S$ in our model are: 
\begin{equation}
\Delta T=\frac{3g_{\rho q\overline{q}}^{2}m_{t}^{2}}{32\pi ^{2}\alpha
_{em}\left( m_{Z}\right) \left( g^{2}+g_{\rho q\overline{q}}^{2}\right) v^{2}%
}+\frac{3}{16\pi \cos ^{2}\theta _{W}}\left[ \frac{\alpha _{2}g_{\rho
}g_{\rho q\overline{q}}+g_{\rho q\overline{q}}^{2}}{g^{2}+g_{\rho q\overline{%
q}}^{2}}-\frac{\alpha _{2}^{2}g_{\rho }^{2}g_{\rho q\overline{q}}^{2}}{%
2g^{2}\left( g^{2}+g_{\rho q\overline{q}}^{2}\right) }\right] \ln \left( 
\frac{m_{h}^{2}}{m_{W}^{2}}\right) ,  \label{DeltaT}
\end{equation}%
\begin{eqnarray}
\Delta S &=&\frac{1}{12\pi }\left( \frac{g-\sqrt{g^{2}+g_{\rho q\overline{q}%
}^{2}}}{\sqrt{g^{2}+g_{\rho q\overline{q}}^{2}}}-\frac{\alpha _{2}g_{\rho
}g_{\rho q\overline{q}}}{2g\sqrt{g^{2}+g_{\rho q\overline{q}}^{2}}}\right)
\ln \left( \frac{m_{h}^{2}}{m_{W}^{2}}\right) +\frac{1}{2\pi }\left[ 3-\frac{%
1}{3}\ln \left( \frac{m_{t}^{2}}{m_{b}^{2}}\right) \right] \left( \frac{%
g_{\rho q\overline{q}}^{2}}{g\sqrt{g^{2}+g_{\rho q\overline{q}}^{2}}}+\frac{%
g-\sqrt{g^{2}+g_{\rho q\overline{q}}^{2}}}{\sqrt{g^{2}+g_{\rho q\overline{q}%
}^{2}}}\right)  \notag \\
&&+\frac{\cos ^{2}\theta _{W}\cos 2\theta _{W}}{3\pi }\left\{ \frac{\Lambda
^{2}}{M_{\rho }^{2}}-\frac{27}{8}-\frac{29}{4}\ln \left( \frac{\Lambda
^{2}+M_{\rho }^{2}}{M_{\rho }^{2}}\right) +\frac{\left( 34M_{\rho
}^{2}\Lambda ^{2}+27M_{\rho }^{4}\right) }{8\left( \Lambda ^{2}+M_{\rho
}^{2}\right) ^{2}}+\frac{35\Lambda ^{2}}{4\left( \Lambda ^{2}+M_{\rho
}^{2}\right) }\right\}  \notag \\
&&\times \left( \frac{g}{\sqrt{g^{2}+g_{\rho q\overline{q}}^{2}}}+\frac{%
g_{\rho }g_{\rho q\overline{q}}}{g\sqrt{g^{2}+g_{\rho q\overline{q}}^{2}}}%
\right).  \label{DeltaS}
\end{eqnarray}%
\newline
%\par
%\vspace{-0.3cm}
Let us note that in our framework of strongly interacting effective theory, the $S$ parameter does not get tree level contributions because there are no
kinetic mixing terms between the gauge fields and the heavy vectors. Here
all couplings of the gauge fields with the heavy vectors are given by the
mass terms. 
%Let us note that in strongly interacting models having kinetic mixingsbetween heavy vector resonances and the SM gauge fields, the $S$ parameter has a tree level contribution. This situation is not present in our model since it does not include a kinetic mixing between the SM gauge fields and the heavy vectors. In our model, the SM gauge bosons and the heavy vectorsonly mix via Lagrangian mass terms as indicated by Eq. (\ref{ModelLagrangian}). 
Using the expressions given above, it follows that for the benchmark point $%
\alpha _{2}=$ $g_{\rho }=1$ and $g_{\rho q\overline{q}}=0.14$ (so that the $%
Zb\bar{b}$ constraint is fullfilled), we get $\Delta T\simeq 4.24\times
10^{-2}$ and $-0.13\lesssim \Delta S\lesssim -0.08$ for heavy vector masses
in the range $2.2$ TeV $\lesssim M_{\rho }\lesssim $ $3$ TeV. On the other
hand, for the benchmark point $\alpha _{2}=$ $g_{\rho }=3.5\lesssim \sqrt{%
4\pi }$ and $g_{\rho q\overline{q}}=0.14$, we find $\Delta T\simeq -0.15$
and $-0.24\lesssim \Delta S\lesssim -0.15$, also for heavy vector masses in
the range $2.2$~TeV $\lesssim M_{\rho }\lesssim $ $3$ TeV.

\begin{figure*}[tbh]
\resizebox{0.7\textwidth}{!}{
\includegraphics{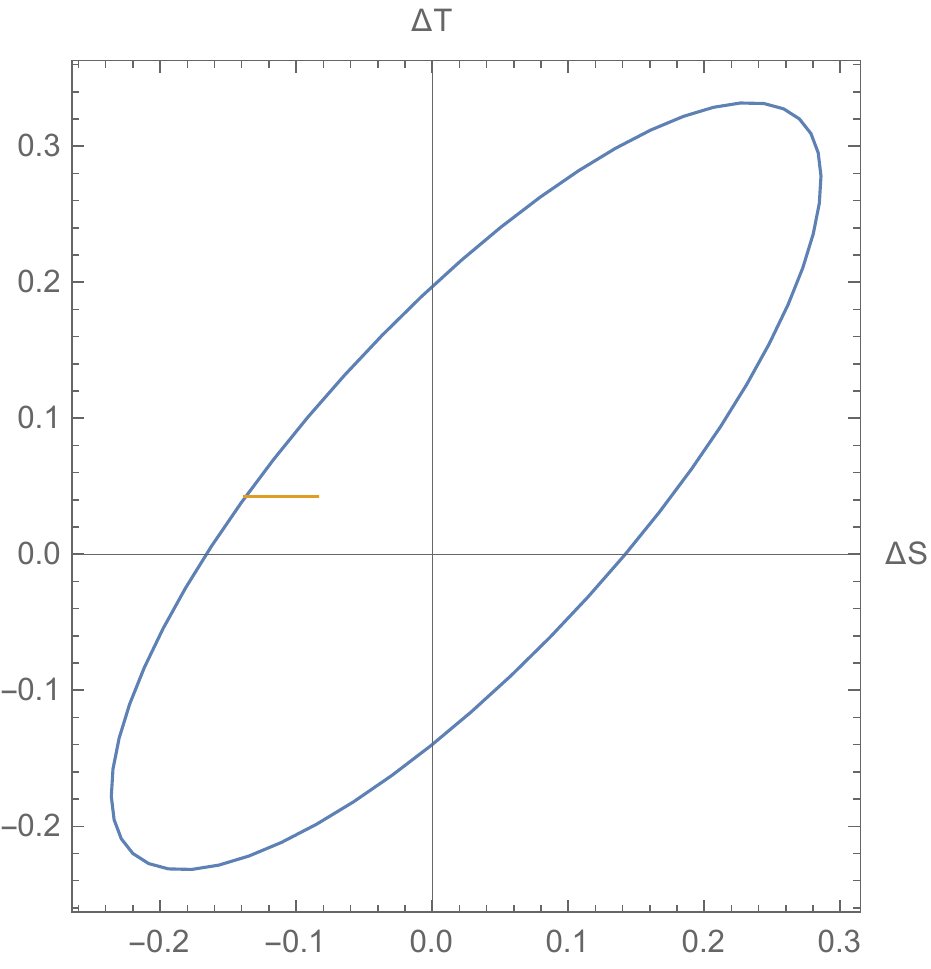}\hspace{2cm}\includegraphics{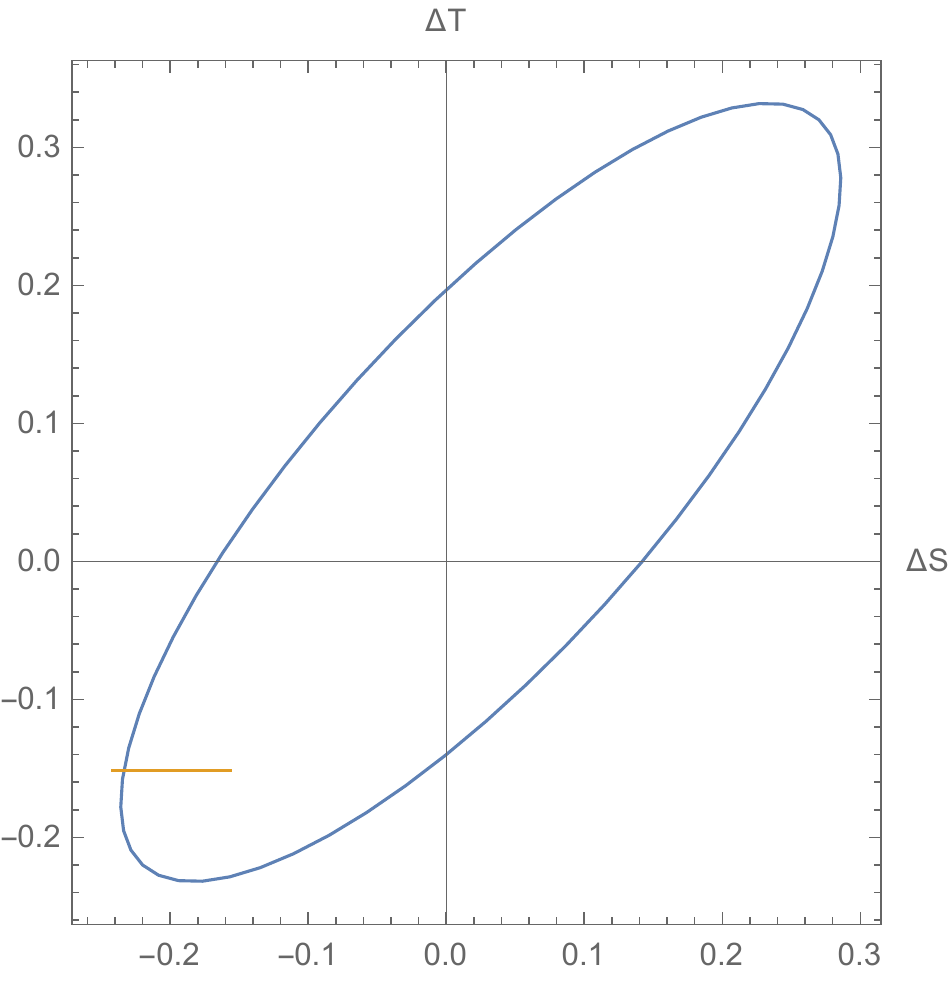}}\newpage
\hspace{2cm}{\footnotesize {{(\ref{fig1}.a)} $\alpha _{2}=$ $g_{\rho }=1$, $g_{\rho q\overline{q}}=0.14$}}\hspace{2cm}%
{\footnotesize{(\ref{fig1}.b)} $\alpha _{2}=$ $g_{\rho }=3.5$, $g_{\rho q\overline{q}}=0.14$%
}\hspace{5.5cm}\newline
%{\footnotesize {\hspace{15cm} \ \ }\hspace{6cm}{}\newline
\caption{The $\Delta S-\Delta T$ plane in our model. The ellipses denote the
experimentally allowed region at $95\%$CL taken from \protect\cite{GFitter}.
The origin $\Delta S=\Delta T=0$ corresponds to the Standard Model value,
with $m_{h}=126$~GeV and $m_{t}=176$~GeV. Figures a and b correspond to
three different sets of values for the couplings $\protect\alpha _{2}$, $g_{%
\protect\rho }$, as indicated. The horizontal line shows the values of $%
\Delta S$ and $\Delta T$ in the model, as the mass of the heavy vectors $M_{%
\protect\rho}$ varies over the range $2.2$ TeV $\leq M_{\protect\rho}\leq$ $3
$ TeV.}
\label{TandSmodel}
\end{figure*}

In Figs.\ \ref{TandSmodel}.a and \ref{TandSmodel}.b we show the allowed
regions for the $\Delta T$ and $\Delta S$ parameters, for the two sets of
values of $\alpha _{2}$, $g_{\rho }$ and $g_{\rho q\overline{q}}$ previously
indicated. The ellipses denote the experimentally allowed region at 95\%
C.L., while the horizontal line shows the values of $\Delta T$ and $\Delta S$
in the model, as the mass of the heavy vectors $M_{\rho}$ is varied from $2.2
$ TeV up to $3$ TeV. 
%The lines are horizontal because $\Delta T$ does not depend on $M_{\rho}$. 
%Consequently 
As shown,  
%the obtained values for 
the $\Delta T$ and $\Delta S$ parameters in our model stay inside the 
% experimentally allowed region at $95\%$ C.L, corresponding to the interior of the 
ellipse  in Fig.~\ref{TandSexp} for almost all the region of parameter space, and consequently 
%. In other words, 
the $T$ and $S$ constraints are easily fulfilled within our
model, 
%and therefore do not impose 
not imposing any restrictions on the model parameters other than the assumption that the heavy vector masses lie in the
range $2.2$ TeV $\lesssim M_{\rho }\lesssim $ $3$ TeV. The lower bound comes
from the ATLAS lower bound of $2.2$ TeV on direct searches for dijet
resonances, while the upper bound is simply the assumed compositeness scale $%
\Lambda \sim 3$ TeV. 
Besides, the line in the figure is practically horizontal 
% Let us note that the dominant contribution to the 
% $\Delta T$ parameter does not depend on the heavy vector masses, 
because 
%since 
the contribution of the heavy vectors to the $\Delta T$ parameter is very small.
The natural smallness of the heavy vector contributions to the $\Delta T$ in
our model is due to its form $\Delta T\sim v^{2}/M_{\rho }^{2}$, where the
heavy vector masses $M_{\rho }$ are clearly much larger than the vacuum
expectation value $v$.

Concerning $\Delta S$, even though its expression given in Eq. \ref{DeltaS}
exhibits a quadratic divergence with the cutoff, in the range of parameters
of our model, the numerical values for $\Delta S$ are well within the
experimentally allowed range. Indeed, Fig ~\ref{DeltaSvsLambda} shows the
sensitivity of the $\Delta S$ parameter under variations of the cutoff $%
\Lambda $ for different values of the heavy vector masses $M_{\rho }$,
namelly, $2.2$ TeV, $2.5$ TeV and $2.8$ TeV. Here we vary the cutoff $\Lambda $ 
% of our model 
 from $3$ TeV to $5$ TeV. As can be seen from Fig ~%
\ref{DeltaSvsLambda}, the $\Delta S$ parameter decreases when the cutoff is
increased and from some values of the cutoff (depending on the heavy vector
masses), it goes outside the allowed experimental limits but having the same
order of magnitude. Consequently, one can say that the variation of the $%
\Delta S$ parameter with the cutoff is rather weak, thus making the $\Delta S
$ parameter acquiring values of the same order of magnitude than the allowed
experimental limits.

\section{Higgs diphoton rate and Higgs boson mass.}

\label{Higgsdiphotonrate} 
\begin{figure*}[tbh]
\begin{center}
\resizebox{0.8\textwidth}{!}{
\includegraphics{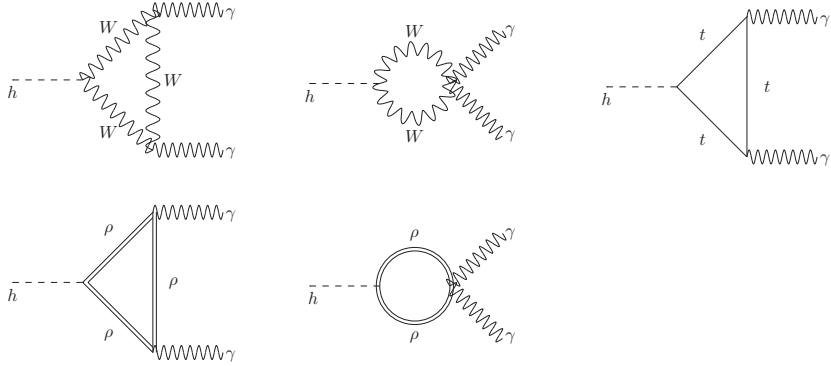}}
\end{center}
\par
\vspace{-12.5cm}
\caption{One loop Feynman diagrams in the Unitary Gauge contributing to the $%
h\rightarrow \protect\gamma \protect\gamma $ decay.}
\label{hto2photons}
\end{figure*}

In the Standard Model, the $h\rightarrow \gamma
\gamma $ decay mode is dominated by $W$ loop diagrams which can interfere
destructively with the subdominant top quark loop. In our strongly coupled
model, the $h\rightarrow \gamma \gamma $ decay receives additional
contributions from loops with charged $\rho _{\mu }^{\pm }$, as shown in
Fig.~\ref{hto2photons}. The explicit form for the $h\rightarrow \gamma
\gamma $ decay rate is: 
\begin{equation}
\Gamma \left( h\rightarrow \gamma \gamma \right) =\frac{\alpha
_{em}^{2}\,m_{h}^{3}}{256\pi ^{3}v^{2}}\left\vert
\sum_{f}N_{c}Q_{f}^{2}F_{1/2}\left( x_{f}\right) +F_{1}\left( x_{W}\right)
+a_{h\rho ^{+}\rho ^{-}}F_{1}\left( x_{\rho }\right) \right\vert ^{2},
\end{equation}%
where: 
\begin{equation}
a_{h\rho ^{+}\rho ^{-}}\simeq\frac{g_{C}^{2}v^{2}}{M_{\rho ^{\pm }}^{2}},%
\hspace{1.5cm}g_{C}^{2}=\left( \beta _{1}+\frac{\beta _{2}}{2}\right) \frac{%
g_{\rho }^{2}}{2}.  \label{coupling1}
\end{equation}%
Here $x_{i}$ are the mass ratios $x_{i}=m_{h}^{2}/4M_{i}^{2}$, with $%
M_{i}=m_{f},M_{W}$ or $M_{\rho }$, respectively, $\alpha _{em}$ is the fine
structure constant, $N_{C}$ is the color factor ($N_{C}=1$ for leptons, $%
N_{C}=3$ for quarks), and $Q_{f}$ is the electric charge of the fermion in
the loop. From the fermion loop contributions we will keep only the dominant
term, which is the one involving the top quark.

The dimensionless loop factors $F_{1/2}\left( x\right) $ and $%
F_{1}\left(x\right) $ (for particles of spin $1/2$ and 1 in the loop,
respectively) are \cite%
{Ellis:1975ap,HppBorn,HppBorn0,HppAnnecy,Gunion:1989we,Spira:1997dg,Djouadi2008,Marciano:2011gm}%
: 
\begin{equation}
F_{1/2}\left( x\right) =2\left[ x+\left( x-1\right) f\left( x\right) \right]
x^{-2},
\end{equation}%
\begin{equation}
F_{1}\left( x\right) =-\left[ 2x^{2}+3x+3\left( 2x-1\right) f\left( x\right) %
\right] x^{-2},  \label{F}
\end{equation}%
with 
\begin{equation}
f\left( x\right) =%
\begin{cases}
\arcsin ^{2}\sqrt{x},\hspace{0.5cm}\mathit{for}\hspace{0.2cm}x\leq 1 \\ 
-\frac{1}{4}\left[ \ln \left( \frac{1+\sqrt{1-x^{-1}}}{1-\sqrt{1-x^{-1}}}%
\right) -i\pi \right] ^{2},\hspace{0.5cm}\mathit{for}\hspace{0.2cm}x>1.%
\end{cases}%
\end{equation}
In what follows, we want to determine the range of values for the mass $%
M_{\rho}$ of the heavy vector resonances, which is consistent with the Higgs
diphoton signal strength measured by the ATLAS and CMS collaborations at the
LHC. %with the $h\rightarrow \gamma \gamma $ results at the LHC. 
To this end, we introduce the ratio $R_{\gamma \gamma }$, which corresponds
to the Higgs diphoton signal strength that normalises the $\gamma\gamma$
signal predicted by our model relative to that of the SM: 
\begin{eqnarray}
R_{\gamma \gamma }&=&\frac{\sigma\left(pp\rightarrow h \right)\Gamma \left(
h\rightarrow \gamma \gamma \right) }{\sigma\left(pp\rightarrow h
\right)_{SM}\Gamma \left( h\rightarrow \gamma \gamma \right) _{SM}}  \notag
\\
&\simeq&\frac{\Gamma \left( h\rightarrow \gamma \gamma \right) }{\Gamma
\left( h\rightarrow \gamma \gamma \right) _{SM}}.  \label{R_gamma}
\end{eqnarray}

\begin{figure*}[tbh]
\begin{center}
\vspace{1cm} \resizebox{0.6\textwidth}{!}{\includegraphics{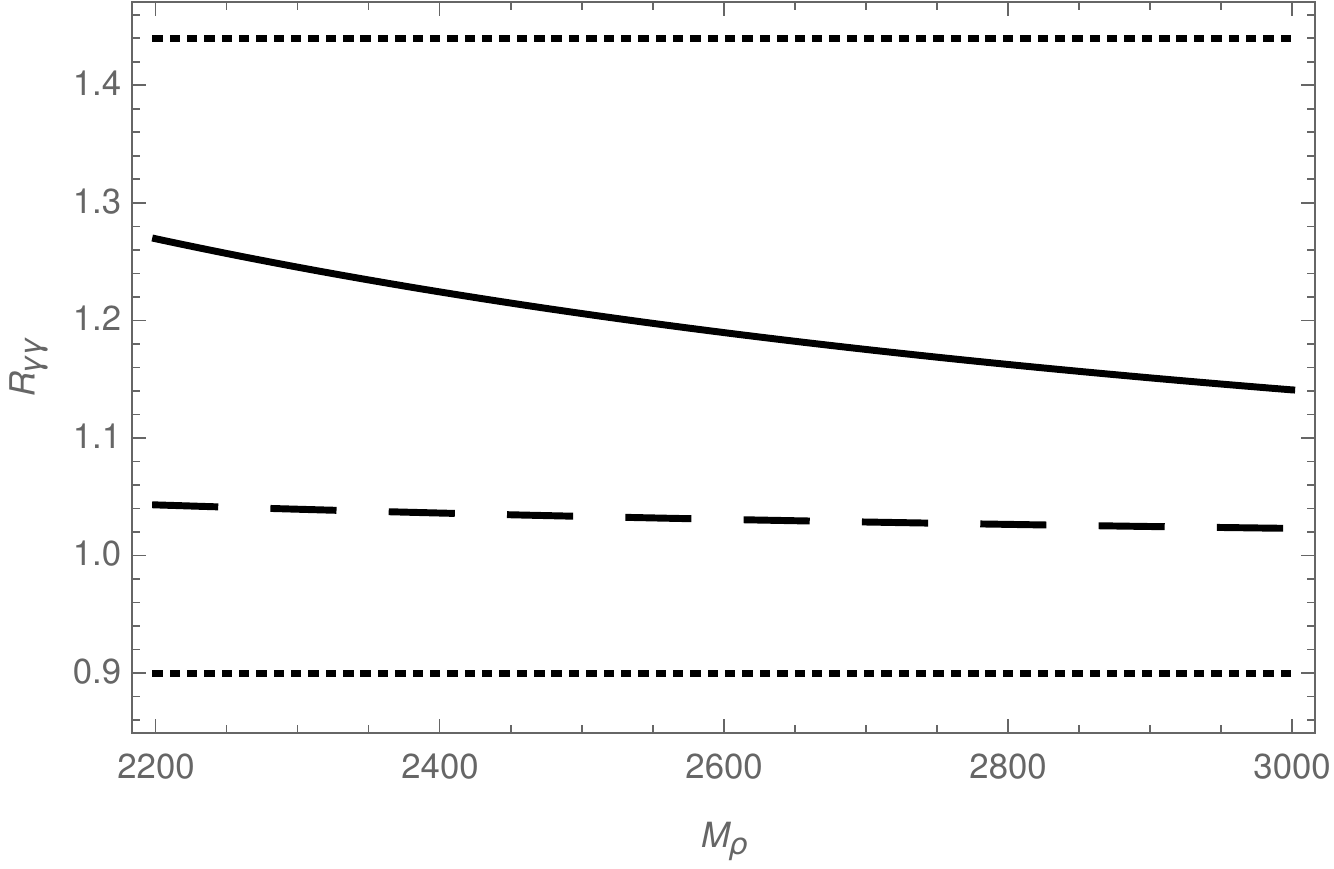}}
\end{center}
\par
%\vspace{-0.3cm}
\caption{The Higgs diphoton signal strength $R_{\protect\gamma \protect\gamma }$ as a function of the mass $M_{\protect\rho }$ of the heavy vectors
for $g_C=3.07$ (solid curve) and $g_C=1.26$ (dashed curve). Here the values $g_C=3.07$ and $g_C=1.26$ correspond to the choices $\beta_1=\beta_2=1$, $g_{\rho}=\sqrt{4\pi}\simeq 3.5$ (maximum value allowed by perturbativity) and $\beta_1=\beta_2=1$, $g_{\rho}=1.45$, respectively (see Eq. \ref{coupling1} ). The horizontal lines
are minimum and maximum values of the ratio $R_{\protect\gamma\protect\gamma }$ inside the 1$\protect\sigma$ experimentally allowed range, determined by
the experimental values and their uncertainties given by CMS and ATLAS, which are equal to $1.14^{+0.26}_{-0.23}$ and $1.17\pm 0.27$, respectively 
\protect\cite{CMS2014,ATLAS2014}. Here we used the values given by ATLAS, which span a broader uncertainty range  than CMS.}
\label{fig1}
\end{figure*}

This normalization for $h\rightarrow \gamma \gamma$ was also done in Refs.~%
\cite{Wang,Campos:2014zaa,Hernandez:2015dga}. Here we have used the fact that in our model,
single Higgs production is also dominated by gluon fusion as in the Standard
Model. 
Fig.~\ref{fig1} shows the sensitivity of the ratio $R_{\gamma \gamma }$
under variations of the heavy vector masses $M_{\rho }$ for different values
of the effective coupling $g_{C}$ (Figs. \ref{fig1}.a and \ref{fig1}.b
correspond to $g_{C}=3.07$ and $g_{C}=1.26$, respectively). As previously
mentioned, we only consider heavy vector masses above the ATLAS lower bound
of $2.2$ TeV for dijet measurements, and up to the compositeness cutoff $%
\Lambda \sim 3$ TeV of our model. We see that an increase of the effective
coupling $g_{C}$, which will correspond to a strong coupling of the heavy
resonances with the Higgs boson, will give rise to an excess of events in
the Higgs diphoton decay channel when compared with the SM expectation. In
that case the Higgs diphoton signal strength will decrease from $1.27$ up to 
$1.14$ when the heavy vector masses are increased from $2.2$ TeV up to $3$
TeV, as indicated by Fig. \ref{fig1}.a. Requiring that the Higgs diphoton
signal strength stays in the ballpark $0.9\lesssim R_{\gamma \gamma
}\lesssim 1.44$ (the value obtained when we use the experimental errors of
the recent CMS and ATLAS results, respectively), we find that heavy vector
masses in the range $2.2$ TeV $\lesssim M_{\rho }\lesssim $ $3$ TeV are
consistent with this requirement. 
%the number of events of the Higgs diphoton decay observed at the LHC. 
On the other hand, as the effective coupling $g_{C}$ gets smaller, which
implies that the coupling of the heavy resonances with the Higgs boson
becomes weaker, the effect of these composite vector resonances in the Higgs
diphoton decay rate turns out to be negligible, giving rise to a Higgs
diphoton decay rate very close to that one predicted by the Standard Model,
as indicated by Fig. \ref{fig1}.b. 
%lead to large values of the compo  %We see that the Higgs diphoton decay rate is very close to the predicted by the Standard Model. 

\begin{figure*}[tbh]
\begin{center}
\resizebox{0.95\textwidth}{!}{
\includegraphics{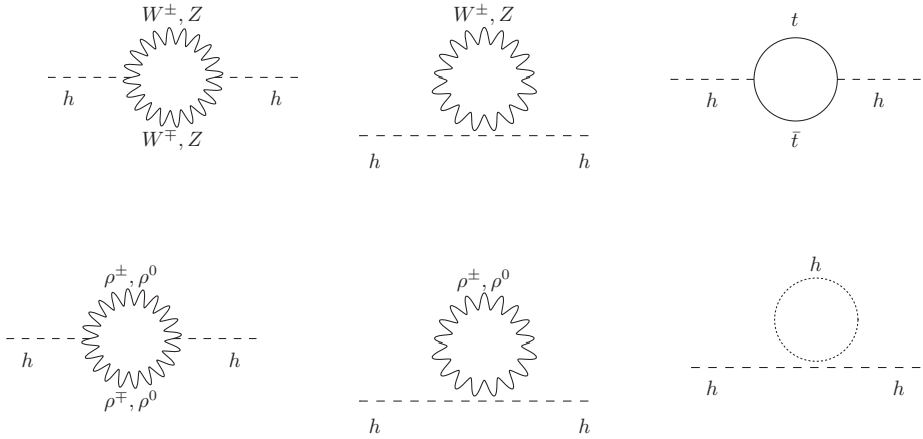}}\vspace{-15cm}
\end{center}
\caption{One loop Feynman diagrams in the Unitary Gauge contributing to the
Higgs boson mass.}
\label{Diagmh}
\end{figure*}

\begin{figure*}[tbh]
\begin{center}
\vspace{0.5cm} 
\resizebox{0.6\textwidth}{!}{
\includegraphics{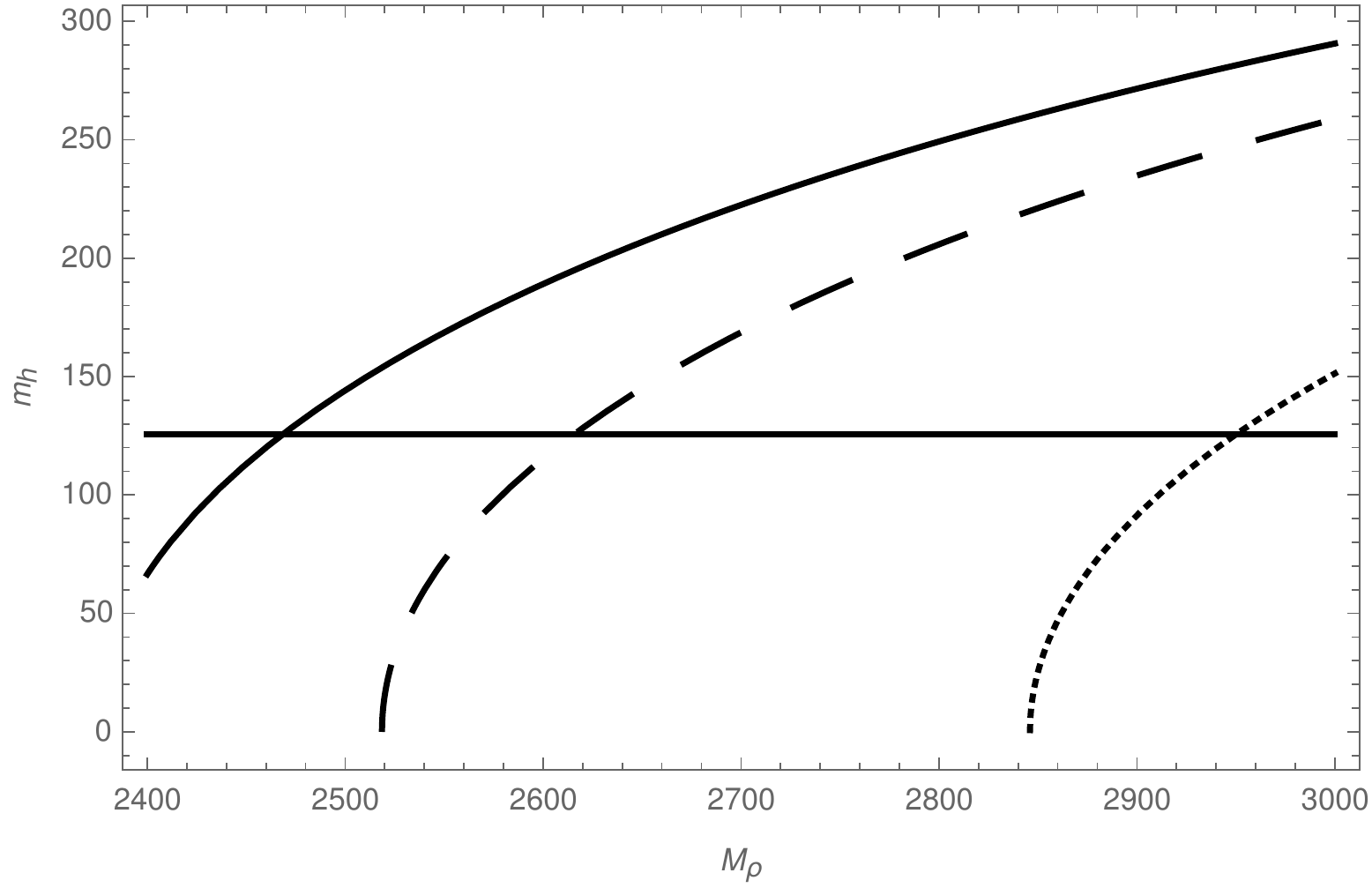}}
\end{center}
\caption{Higgs boson mass $m_{h}$ as function of the heavy vector mass $M_{%
\protect\rho }$ for different values of the effective coupling $g_{C}$. The
horizontal line corresponds to the value 126 GeV for the light Higgs boson
mass. The solid, dashed and dotted curves correspond to the cases where $g_{C}$
is set to be equal to $1.20$, $1.26$ and $1.40$, respectively. The quartic
Higgs coupling is taken to be equal to that of the Standard Model. Here we fix the tree level Higgs boson mass  $\left(m_{h}\right) _{0}$ to be equal to $126$ GeV.}
\label{fig2}
\end{figure*}

\begin{figure}[tbh]
\begin{center}
\resizebox{0.45\textwidth}{!} {\includegraphics{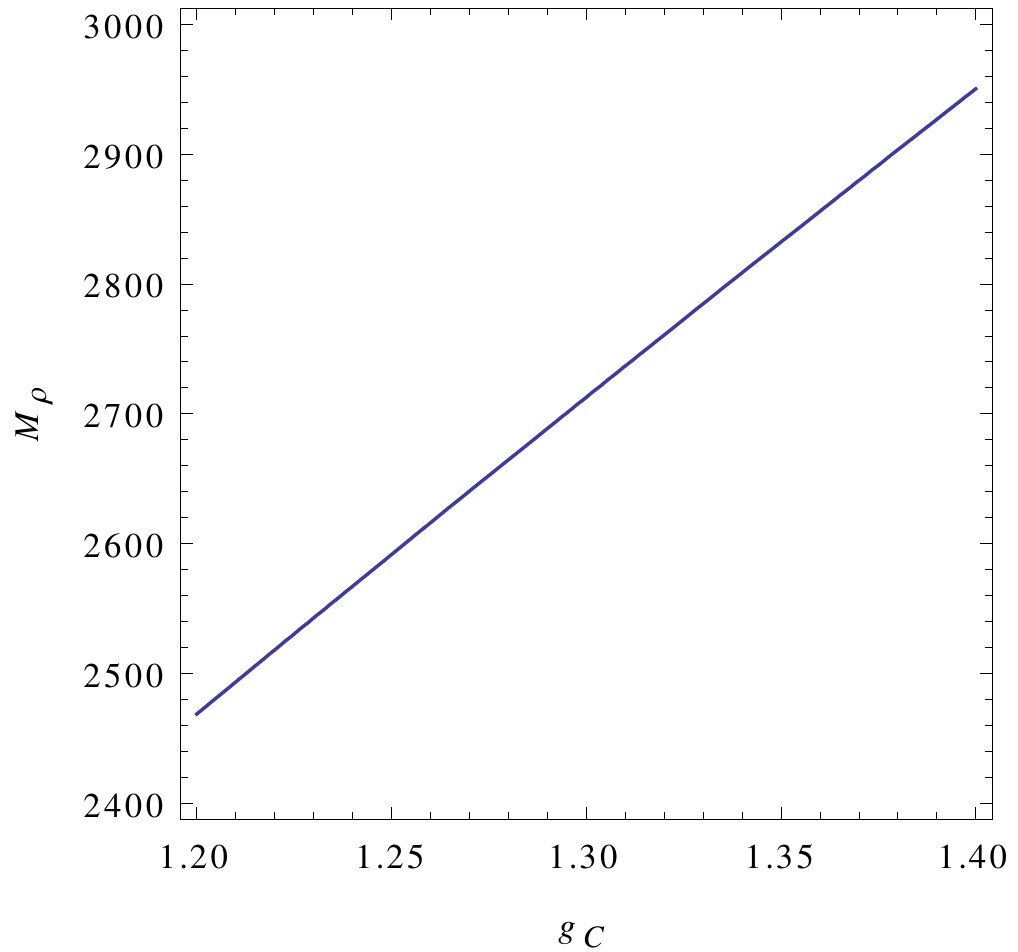}}
\end{center}
\caption{Correlation between the $g_{C}$ effective coupling and the heavy vector mass $M_{\protect\rho }$ consistent with a Higgs boson mass of $126$ GeV. The quartic Higgs coupling is taken to be equal to that of the Standard Model.}
\label{fig3}
\end{figure}

Let us now determine the contraints on the effective coupling $g_{C}$ and
the masses $M_{\rho }$ of the heavy vector resonances that can successfully
accommodate a $126$ GeV Higgs boson mass. To this end, we proceed to compute
the tree level and one loop level contributions to the Higgs boson mass $%
m_{h}$. 
%The Higgs boson mass $m_{h}$ receives contributions at tree and at one loop level corrections. 
The squared Higgs boson mass is given by:%
\begin{equation}
m_{h}^{2}=\left( m_{h}^{2}\right) _{0}+\Sigma _{h},
\end{equation}%
where $\left( m_{h}^{2}\right) _{0}$ is the squared tree level Higgs boson mass
and $\Sigma _{h}$ corresponds to the one loop level contribution, arising
from Feynman diagrams containing spin-$0$, spin-$1/2$ and spin-$1$ particles
in the internal lines of the loops. For the contribution from the fermion
loops we will only keep the dominant term, which is the one involving the
top quark. From the Feynman diagrams shown in Figure \ref{Diagmh}, it
follows that the one loop level contribution to the squared Higgs boson mass
is given by: 
\begin{eqnarray}
\Sigma _{h} &\simeq &2F_{A}\left( M_{W}\right) +2F_{B}\left( M_{W}\right)
+F_{A}\left( M_{Z}\right) +F_{B}\left( M_{Z}\right)  \notag \\
&&+2a_{h\rho ^{+}\rho ^{-}}^{2}F_{A}\left( M_{\rho }\right) +2a_{h\rho
^{+}\rho ^{-}}F_{B}\left( M_{\rho }\right)  \notag \\
&&+a_{h\rho ^{0}\rho ^{0}}^{2}F_{A}\left( M_{\rho }\right) +a_{h\rho
^{0}\rho ^{0}}F_{B}\left( M_{\rho }\right)  \notag \\
&&+F_{C}\left( m_{t}\right) +\frac{3}{4}\lambda F_{D}\left( m_{h}\right)
\label{mhloop}
\end{eqnarray}%
where $a_{h\rho ^{0}\rho ^{0}}\simeq a_{h\rho ^{+}\rho ^{-}}$ with the
dimensionless parameter $a_{h\rho ^{+}\rho ^{-}}$ given by Eq. \ref%
{coupling1}. In addition, the loop functions appearing in Eq. \ref{mhloop}
are: 
\begin{eqnarray}
F_{A}\left( M\right) &=&\frac{M^{4}}{32\pi ^{2}v^{2}}\left[ \frac{\Lambda
^{4}}{M^{4}}-\frac{6\Lambda ^{2}}{M^{2}+\Lambda ^{2}}+6\ln \left( \frac{%
\Lambda ^{2}+M^{2}}{M^{2}}\right) \right] ,  \notag \\
F_{B}\left( M\right) &=&-\frac{M^{2}}{32\pi ^{2}v^{2}}\left[ \frac{\Lambda
^{4}}{M^{2}}+6\Lambda ^{2}-6M^{2}\ln \left( \frac{\Lambda ^{2}+M^{2}}{M^{2}}%
\right) \right] ,  \notag \\
F_{C}\left( M\right) &=&\frac{3M^{2}}{4\pi ^{2}v^{2}}\left[ \allowbreak
\Lambda ^{2}-3M^{2}\ln \left( \frac{\Lambda ^{2}+M^{2}}{M^{2}}\right) +\frac{%
3M^{2}\Lambda ^{2}}{\Lambda ^{2}+M^{2}}\right] ,  \notag \\
F_{D}\left( M\right) &=&-\frac{1}{16\pi ^{2}}\left[ \Lambda ^{2}-M^{2}\ln
\left( \frac{\Lambda ^{2}+M^{2}}{M^{2}}\right) \right] .  \label{Floop}
\end{eqnarray}

In Fig. \ref{fig2} we show the sensitivity of the Higgs boson mass $m_{h}$
to variations in $M_{\rho }$ for $g_{C}=1.20$, $1.26$ and $1.40$ and the
quartic Higgs coupling set to be equal to the Standard Model value. 
%different values of the effective coupling $g_C$, set to be equal to $1.20$, $1.26$ and $1.40$. 
%The heavy vector masses are taken to range from $2.2$ TeV and $3$ TeV. 
The Higgs boson mass $m_{h}$ is an increasing function of the heavy vector
masses $M_{\rho }$. 
%, i.e., the Higgs boson mass is increased as the heavy vector masses take larger values. 
These Figures show that the heavy vector masses $M_{\rho }$ have an
important effect of $m_{h}$. This is due to the fact that the one loop
diagrams involving the heavy vector resonances and contributing to the Higgs
boson mass are very sensitive to the cuttoff $\Lambda $, since they exhibit
quartic and quadratic divergences that are not cancelled. Let us note that
these heavy vector resonance one loop contributions to the Higgs boson mass
involve trilinear and contact interactions, whose dominant terms are
proportional to $\frac{g_{C}^{2}\Lambda ^{4}}{M_{\rho }^{2}}$ and $\frac{g_{C}^{4}v^{2}\Lambda ^{4}}{M_{\rho }^{4}}$, respectively, and consequently
for larger vector masses, a stronger effective $g_{C}$ coupling is required
to reproduce the $126$ GeV value for the Higgs boson mass, as shown in Fig. %
\ref{fig3}.

\section{Decay channels of the heavy vectors}

\label{Decaychannels}

%The current important period of LHC exploration of the Higgs boson interactions may provide a crucial step to unravel the electroweak symmetry breaking mechanism, by determining the right explicit model, among the different scenarios based on a strongly interacting dynamics for electroweak symmetry. Because of this reason, we consider relevant to determine the smoking gun that allows to directly test our model in the second run of the LHC, by studying the most relevant decay channels of the heavy vector resonances that could constitute direct signatures at the LHC. 
The current important period of LHC exploration of the Higgs properties and
discovery of heavier particles may provide crucial steps to unravel the
electroweak symmetry breaking mechanism. Consequently, we complement our
work by studying the most relevant decay channels of the heavy vector
resonances that could constitute direct signatures of our scenario at the
LHC. To this end, we compute the two body decay widths of the heavy vectors.
These widths, up to corrections of order $m_{h}^{2}/M_{\rho }^{2}$ and $%
M_{W}^{2}/M_{\rho }^{2}$ are: 
\begin{eqnarray}
\Gamma \left( \rho ^{0}\rightarrow q\overline{q}\right) &\simeq &\frac{%
3g_{\rho q\overline{q}}^{2}}{96\pi }M_{\rho },  \notag \\
\Gamma \left( \rho ^{+}\rightarrow u_{i}\overline{d}_{j}\right) &=&\Gamma
\left( \rho ^{-}\rightarrow \overline{u}_{i}d_{j}\right) \simeq \frac{%
3g_{\rho q\overline{q}}^{2}}{96\pi }\left\vert V_{ij}\right\vert ^{2}M_{\rho
},  \notag \\
\Gamma \left( \rho ^{\pm }\rightarrow W^{\pm }h\right) &=&\Gamma \left( \rho
^{0}\rightarrow Zh\right) \simeq \frac{\alpha _{2}^{2}g_{\rho }^{2}}{384\pi }%
M_{\rho },  \notag \\
\Gamma \left( \rho ^{0}\rightarrow W^{+}W^{-}\right) &=&\Gamma \left( \rho
^{\pm }\rightarrow W^{\pm }Z\right) \simeq \frac{\alpha _{2}^{2}g_{\rho }^{2}%
}{384\pi }M_{\rho }.  \label{decaymodes}
\end{eqnarray}

\begin{figure*}[tbh]
\resizebox{17cm}{6cm}%{\rule{6cm}{7cm}}
%{1\textwidt}{!}
{\includegraphics{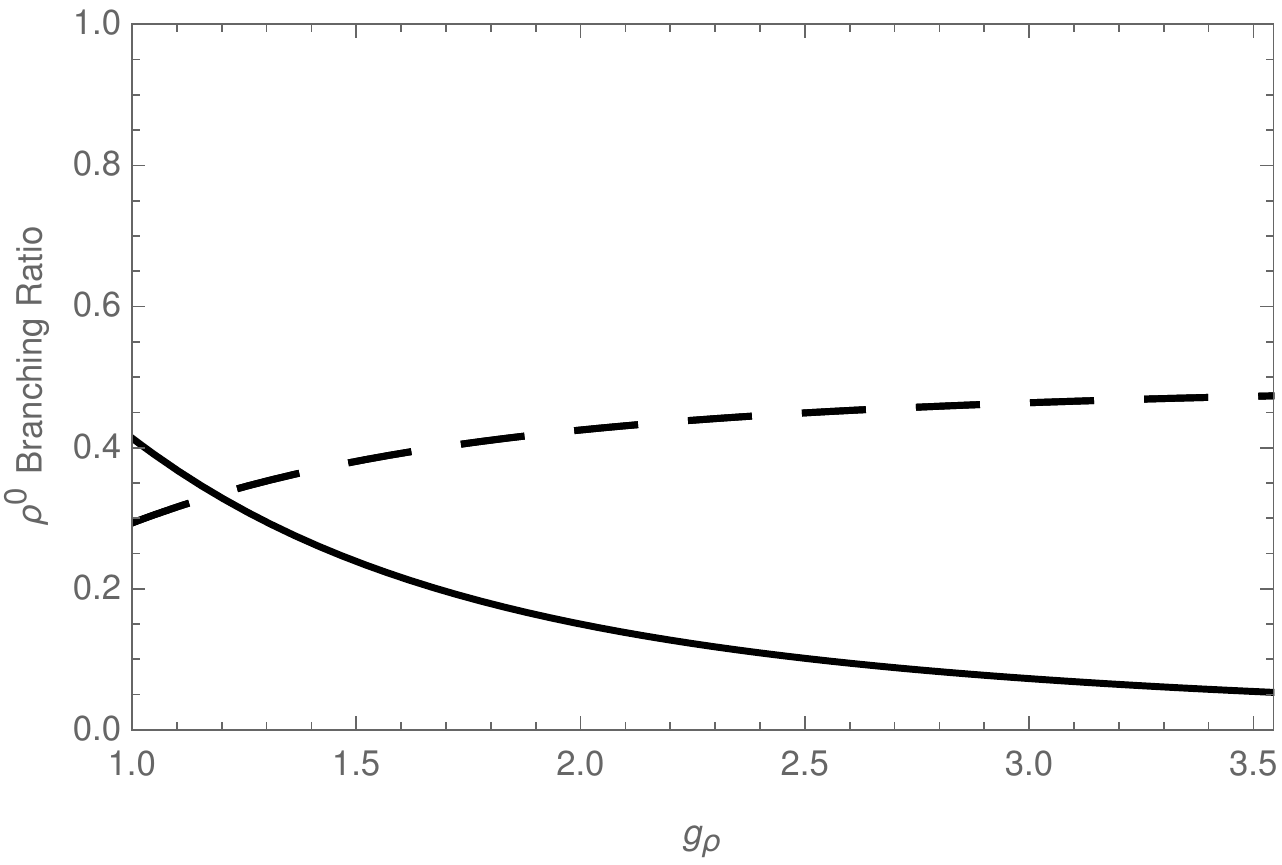} \includegraphics{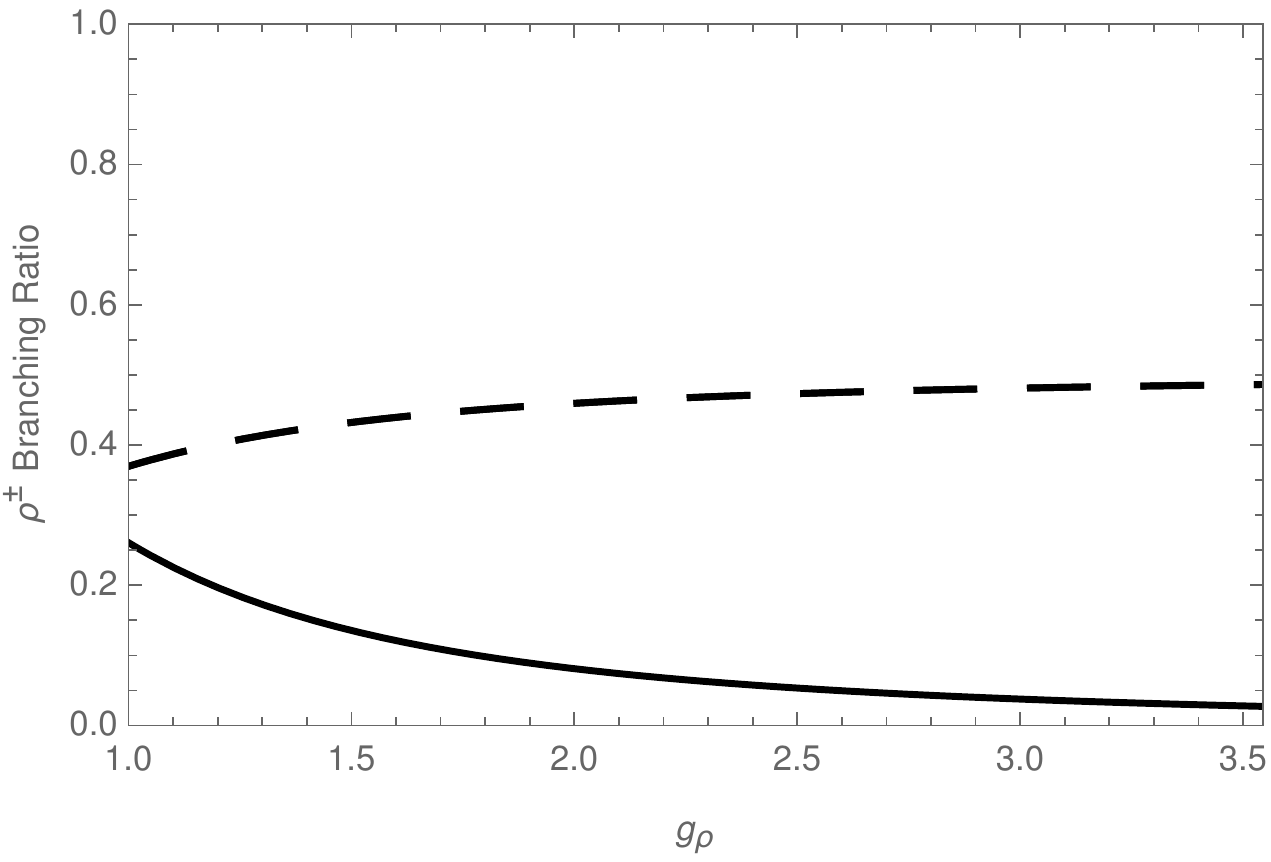}} 
\begin{tabular}{cccc}
\hspace{3cm} & (\ref{BR}.a) & \hspace{8cm} & (\ref{BR}.b)%
\end{tabular}%
\caption{Branching ratios of the neutral (Fig.\ \protect\ref{BR}.a) and charged (Fig.\ \protect\ref{BR}.b) heavy vectors as functions of $g_{\protect%
\rho}$ for $\protect\alpha_2=1$ and $g_{\protect\rho q\overline{q}}=0.14$. The solid curves correspond to the Branching ratios of the heavy vectors into quark pairs: a) $6Br\left(\rho ^{0}\rightarrow q\overline{q}\right)$ 
%(Fig.\ \protect\ref{BR}.a), 
and b) $3Br\left(\rho^{+}\rightarrow u\overline{d}\right)=3Br\left(\rho^{-}\rightarrow d\overline{u}\right)$,  
%(Fig.\ \protect\ref{BR}.b), 
whereas the dashed curves correspond to the Branching ratios of the heavy vectors into a SM  boson pair: a) 
 $Br\left(\rho^{0}\rightarrow W^{+}W^{-}\right)=Br\left( \rho^{0}\rightarrow Zh\right)$ 
 %(Fig.\ \protect\ref{BR}.a), 
 and b) $Br\left(\rho^{\pm}\rightarrow W^{\pm}Z\right)=Br\left( \rho ^{\pm }\rightarrow W^{\pm }h\right)$. 
 %(Fig.\ \protect\ref{BR}.b)
 }
\label{BR}
\end{figure*}

Figure \ref{BR} displays the branching rations of the neutral (Fig.\ \ref{BR}%
.a) and charged (Fig.\ \ref{BR}.b) heavy vectors to quark-antiquark pairs
and to a SM-like Higgs in association with a SM gauge boson, as a function
of the composite vector resonance coupling $g_{\rho }$. This coupling is
taken to range from $1$ to $3.5$ (value slightly lower than the maximum
value $\sqrt{4\pi }$ allowed by perturbativity). Here we set $\alpha _{2}=1$
whereas the direct coupling of the heavy vector resonances with quarks 
%heavy  vector-quark-antiquark coupling 
$g_{\rho q\overline{q}}$ is set to be equal to $0.14$, the maximum value
that keeps $A_{b\bar{b}}$ inside the $1\sigma $ experimentally allowed
range, as described in Section \ref{Zbb}. One can notice that for low values
of the $g_{\rho }$ coupling, the heavy vectors have a dominant decay mode
into quark-antiquark pairs, comparable with the decay into a SM-like Higgs
and SM gauge boson. On the other hand, when the value of the $g_{\rho }$
coupling is increased, the fermionic decay modes of the heavy vectors get
suppressed, and the decay modes into a pair of SM Gauge bosons as well as
into a SM-like Higgs and SM gauge boson become the dominant ones.

\section{Conclusions.}

\label{conclusions}

We studied a framework of strongly interacting dynamics for electroweak
symmetry breaking without fundamental scalars, by means of an effective
theory based on the SM gauge group $SU\left( 2\right) _{L}\times U\left(
1\right) _{Y}$, with a $SU\left( 2\right) _{L}$ triplet of heavy vectors. In
this framework, it is assumed that the strong dynamics responsible for
electroweak symmetry breaking gives rise to a composite 
%$SU\left( 2\right) _{L+R}$ 
triplet of heavy vectors and a composite scalar identified with the $126$
GeV Higgs boson, recently discovered at the LHC. 
% and the SM Goldstone bosons. 
It is assumed that the scalar and the heavy composite vectors are the only
resonances that are lighter than the cufoff $\Lambda \simeq 4\pi v\sim 3$
TeV, so that the interactions among themselves and with the SM particles can
be described by an effective chiral Lagrangian. 
%Furthermore, to reduce the parameter space and to make definite predictions, we restrict the model to the case  where the
The inclusion of the heavy vector resonances in the effective Lagrangian is
done by considering them as gauge vectors of a hidden local symmetry. In
this scenario, we determine the constraints arising from the $Zb\bar{b}$
vertex, from the $T$ and $S$ oblique parameters, and the constraints
resulting from the measured Higgs diphoton decay rate and the Higgs mass of $%
126$ GeV. We found that the $Zb\bar{b}$ constraint at $1\sigma $ implies
that the direct coupling of the heavy vector resonances with quarks should
satisfy the upper bound $g_{\rho q\overline{q}}\lesssim 0.14$, which can be
modified as $g_{\rho q\overline{q}}\lesssim 0.21$ by requiring that the $%
A_{bb}$ parameter that characterizes that constraint is inside the $3\sigma $
experimentally allowed range. Consequently the heavy vector resonances
cannot strongly couple with quarks and therefore we found them to have a
dominant decay mode into a SM-like Higgs and SM gauge boson in the region
where the coupling $g_{\rho }$ of the strong sector is large. However, for
low values of the $g_{\rho }$ coupling, these heavy vectors have a dominant
decay mode into quark-antiquark pairs, comparable with the decays into a SM
like Higgs and SM gauge boson as well as into a SM gauge bosons pair.
Furthermore, we find that our model can easily accommodate the $T$ and $S$
oblique parameter constraints, as well as the Higgs diphoton decay rate
constraints, in the whole relevant region $2.2$ TeV $\lesssim M_{\rho
}\lesssim $ $3$ TeV for the heavy vector masses, for $g_{\rho }<\sqrt{4\pi }$
and $g_{\rho q\overline{q}}\lesssim 0.21$. We considered the heavy vector
masses to be in the range $2.2$ TeV $\lesssim M_{\rho }\lesssim $ $3$ TeV,
since our model is strongly interacting with a cutoff of $\Lambda \sim 3$
TeV, while consistency with the ATLAS dijet measurements yields the lower
bound of $2.2$ TeV for the masses of the heavy vector resonances. In
addition, we found that one loop effects are crucial to successfully
reproduce the $126$ GeV Higgs boson mass. The requirement of having a $126$
GeV Higgs boson constrains the effective coupling $g_{C}$ and the mass $%
M_{\rho }$ of the heavy vectors to be in the ranges $1.2\lesssim
g_{C}\lesssim $ $1.4$ and $2.47$ TeV $\lesssim M_{\rho }\lesssim $ $2.95$
TeV, respectively. In summary, an effective theory of strongly interacting
dynamics for electroweak symmetry breaking, having in its particle spectrum
the SM particles and a composite $SU\left( 2\right) _{L+R}$ triplet of
vector resonances, can successfully accommodate a light $126$ GeV Higgs
boson, provided that the Higgs boson has a moderate but not too large
coupling with heavy composite resonances. This framework is consistent with
electroweak precision tests, Higgs diphoton decay rate constraints and the
constraints arising from the $Zb\bar{b}$ vertex. We have shown that the
current experimental data still allows for Higgs boson strongly coupled to a
composite sector, assumed to include a composite $SU\left( 2\right) _{L+R}$
triplet of vector resonances with a mass below the cutoff $\Lambda $.
Finally, we should briefly comment that the tension between the effective $%
g_{C}$ coupling and the Higgs boson mass $m_{h}$ may be alleviated if the
spectrum below the composite scale includes heavy fermions in addition to
the vectors. Determining the effects of this enriched spectrum on the Higgs
diphoton signal strength, the oblique $T$ and $S$ parameters, the $Zb\bar{b}$
vertex and the Higgs boson mass, requires an additional and careful analysis
that we have left outside the scope of this work.

\subsection*{Acknowledgements}

This work was supported in part by Conicyt (Chile) grant ACT-119 ``Institute
for advanced studies in Science and Technology''. C.D. also received support
from Fondecyt (Chile) grant No.~1130617, and A.Z. from Fondecyt grant
No.~1120346. A.E.C.H was partially supported by Fondecyt (Chile), Grant No.
11130115 and by DGIP internal Grant No. 111458.

\end{document}